\documentclass[review]{elsarticle}
\usepackage{hyperref}
\usepackage[utf8]{inputenc}
\usepackage[T1]{fontenc}
\usepackage{placeins}
\usepackage{amssymb}
\usepackage[version=4]{mhchem}
\usepackage{overpic}
\usepackage{graphics}
\usepackage{multirow}
\usepackage{rotating}
\usepackage{textcomp}
\usepackage{float}              
\usepackage{color}
\usepackage{subcaption}
\usepackage{verbatim}
\usepackage{siunitx}
\usepackage{nomencl}                 	

\journal{Journal of Membrane Science}
\biboptions{sort&compress}

\begin{document}
\sloppy

\begin{frontmatter}

\title{On the transport of \ce{CO2} through humidified facilitated transport membranes}

\author[1]{M. Logemann}
\author[2]{J. Gujt}
\author[1]{T. Harhues}
\author[3,4,5]{T. D. K\"uhne\corref{cor1}}\ead{t.kuehne@hzdr.de}
\author[1,6]{M. Wessling}
\cortext[cor1]{Corresponding author}

\address[1]{RWTH Aachen University, Chemical Process Engineering, Forckenbeckstr. 51, 52074 Aachen, Germany}
\address[2]{Dynamics of Condensed Matter and Institute for Lightweight Design with Hybrid Systems, Chair of Theoretical Chemistry, University of Paderborn, Warburger Str. 100, 33102 Paderborn, Germany}
\address[3]{Center for Advanced Systems Understanding (CASUS), Helmholtz Zentrum Dresden-Rossendorf, Untermarkt 20, D-02826 Görlitz, Germany}
\address[4]{TU Dresden, Institute of Artificial Intelligence, Chair of Computational System Sciences, Nöthnitzer Stra{\ss}e 46
D-01187 Dresden, Germany}
\address[5]{Paderborn Center for Parallel Computing and Center for Sustainable Systems Design, University of Paderborn, Warburger Str. 100, 33102 Paderborn, Germany}
\address[6]{DWI Interactive Materials Research, Forckenbeckstr. 50, 52074 Aachen, Germany}

\begin{abstract}
Membrane-based \ce{CO2} removal from exhaust streams has recently gained much attention as a means of reducing emissions and limiting climate change. Novel membranes for \ce{CO2} removal include so called facilitated transport membranes (FTMs), which offer very high selectivities for \ce{CO2} while maintaining decent permeabilities. Recently, these FTMs have been scaled up from laboratory level to plant-sized pilot modules with promising results. However, the molecular details of \ce{CO2} transport in these has not yet been fully unraveled. In this work, experimental studies were combined with quantum-mechanical \textit{ab initio} molecular dynamics simulations to gain insight into the underlying molecular mechanism of \ce{CO2} permeation through FTMs. Various compositions of polyvinyl alcohol (PVA) as the membrane matrix with polyvinyl amine (PVAm), monoethanolamine (MEA), or 4-amino-1-butanol (BA) as carrier molecules were experimentally tested. Our experiments revealed that water was essential for the \ce{CO2} transport and a transport superposition was achieved with a mixed composition of PVAm and MEA in PVA. Furthermore, sorption measurements with PVA were conducted with humidified \ce{N2} and \ce{CO2} to quantify water sorption-induced swelling and its contribution to the gas uptake. 
\\
As the carbonic acid--amine interaction is assumed to cause transport facilitation, electronic structure-based \textit{ab initio} molecular dynamics simulations were conducted to study the transport of \ce{CO2} in the form of carbonic acid along PVAm polymer chains. In particular, the necessity of local water for transport facilitation was studied at different water contents. The simulations show that transport is fastest in the system with low water content and does not happen in the absence of water. 
\end{abstract}


\end{frontmatter}

\section{Introduction}
Reducing the emission of carbon dioxide (\ce{CO2}) is one of the essential prerequisites for limiting the future effects of climate change. The efficient removal of \ce{CO2} from exhaust streams could be a crucial component \cite{figueroa2008advances}. In this field of research, Merkel et al. \cite{merkel2010power} have shown the advantages of polymeric membrane applications in continuous processes. Facilitated transport membranes (FTMs) are particularly promising for such applications, because they enhance \ce{CO2} transport via a reversible reaction mechanism with fixed-site or mobile carriers \cite{matsuyama1999facilitated, zou2006co2, huang2008carbon,deng2009facilitated, deng2010techno}. This mechanism leads to an improved \ce{CO2} separation with both high permeance and high selectivity. 

Nearly all fixed-site and mobile carriers feature primary or secondary amines that are traditionally used in aqueous reagents for \ce{CO2} and hydrogen sulfide (\ce{H2S}) scrubbing. As a result, the predominant reaction mechanisms have been well studied \cite{kohl1997gas}. As FTMs usually work in a water-swollen state, the reaction mechanisms of the carriers are likely to be similar to those in amine scrubbing. However, FTMs maintain the possibility for continuous \ce{CO2} withdrawal with low energy requirements and without the regeneration of amines \cite{zou2006co2, deng2009facilitated, saeed2015co2}.

The first polymeric \ce{CO2}-selective FTMs were prepared in 1999 by Matsuyama et al. \cite{matsuyama1999facilitated}, who integrated the fixed-site carrier polyethylenimine (PEI) into a polyvinyl alcohol (PVA) membrane matrix. The membrane reached a \ce{CO2}/\ce{N2} selectivity of over 100. In the following years, various research groups combined polymeric matrix materials and carriers to further increase \ce{CO2} permeance while maintaining high selectivity \cite{zou2006co2, huang2008carbon,deng2009facilitated, deng2010techno, cai2008gas, deng2010swelling, uddin2012natural, saeed2015co2, sandru2009high, sandru2010composite, kim2013separation, chen2016high, el2008carbon, el2009parametric, el2009carbon, wu2017so2, mondal2014co2, zhao2013co2, amooghin2013novel, qiao2015preparation, tong2015water, maheswari2017separation}.

PVA is a frequently used FTM matrix material because it is highly compatible with other polymers and serves as a stable and hydrophilic membrane material when cross-linked \cite{matsuyama1999facilitated, zou2006co2,cai2008gas, deng2010techno, deng2010swelling, uddin2012natural, saeed2015co2}. The permeation characteristics and swelling degree of wetted PVA membranes are determined by the applied cross-linking conditions and agents \cite{praptowidodo2005influence, bolto2009crosslinked}. For facilitated transport, cross-linking PVA with an aldehyde, such as formaldehyde (FA) and glutaraldehyde (GA), has been the most common route \cite{figueiredo2009poly, rafiq2016role, tong2017facilitated}.

Other common membrane matrix materials are polyvinyl amine (PVAm) \cite{sandru2009high, sandru2010composite, kim2013separation, chen2016high}, or bio-based materials such as chitosan \cite{el2008carbon, el2009parametric, el2009carbon}. PVAm and chitosan also feature amine functionality and are therefore both matrix materials and fixed-site carriers \cite{deng2009facilitated, deng2010techno, wu2017so2}. Besides these two polymers, other employed fixed-site carriers are PEI \cite{matsuyama1999facilitated, mondal2014co2} and polyallylamine (PAAm) \cite{zou2006co2, cai2008gas, huang2008carbon, zhao2013co2}. Also, mobile carriers such as alkanolamines \cite{amooghin2013novel, qiao2015preparation}, amino acids \cite{zhao2013co2, tong2015water, wu2017so2}, and aliphatic amines \cite{qiao2015preparation, maheswari2017separation} have been introduced into FTMs. Especially the combination of fixed-site and mobile carriers have led to enhanced \ce{CO2} permeabilities. Zou et al. \cite{zou2006co2} reached permeabilities of up to 9700~Barrer along with a \ce{CO2}/\ce{H2} selectivity above 500, which is at least ten times more selective than polymeric membranes without transport facilitation.

While most experiments have been conducted with flat sheet membranes on a lab scale, Sandru et al. \cite{sandru2010composite} tested the first hollow fiber FTMs in 2010. Recently, for both approaches, the focus has shifted from material development toward the scaling up of membrane production and the application of FTMs in pilot plants \cite{he2017pilot, hagg2017pilot, salim2018fabrication, dai2019field, han2019field, chen2020fabrication}.

The underlying transport mechanisms within FTMs, however, have not yet been fully explained, although the materials and processes have been thoroughly studied. The influences of relative humidity and carrier concentrations have yet to be analyzed in molecular detail. Assuming that the dissolved \ce{CO2} locally interacts with the amine group and with water as a carbonic acid ion, to study the transport events of the ion along the polymer chain in a molecular mixture with the dissolved water molecules, suitable simulation methods are needed. To grasp these molecular complexities in the three-component mixture, we combined experimental results coupled with electronic structure-based \textit{ab initio} molecular dynamics (AIMD) simulations to provide further insight into the mechanism of \ce{CO2} transport through humidified FTMs.  

 The transport of \ce{CO2} through pure PVA flat sheet membranes was measured in comparison to FTMs with PVA as matrix material. The FTM compositions were prepared with monoethanolamine (MEA) or 4-amino-1-butanol (BA) as the mobile carrier and/or polyvinyl amine (PVAm) as the fixed-site carrier. Measurements were conducted in a dry atmosphere, as well as in one with a relative humidity above 95\%. Gas sorption measurements were conducted with PVA films to assess the swelling behavior of PVA in different humidities and gas atmospheres.

Quantum-mechanical AIMD simulations were conducted to study the transport mechanism with a fixed-site carrier. The transport of \ce{CO2}, in the form of carbonic acid (\ce{H2CO3}), along PVAm polymer chains was investigated in systems with different water contents. Three wetted states were examined: the dry state, without any water present; a semi-dry state with 1.125 water molecules per primary amine group; and a wet state, with 2.25 water molecules per primary amine group. Such a system, albeit simple compared to the polymers used in the experiments, can provide insight into a water-mediated Grotthuss-like hopping mechanism (see Section \ref{subsub:FTMechanism}) \cite{luduena11}, which we believe is the main transport mechanism across the fixed-site carrier FTMs.

\section{Background}
\subsection{Gas separation in polymeric membranes}
\label{subsec:gassep}
For dense polymeric membranes, the solution-diffusion model (SDM) is commonly applied \cite{wijmans1995solution}. According to the SDM, molecules, driven by the chemical potential, dissolve in the membrane surface on the feed side, diffuse through the membrane, and desorb on the permeate side \cite{wijmans1995solution}. The permeability coefficient $P_{i}$ is the product of the sorption coefficient $S_i$ and the diffusion coefficient $D_i$, i.e. 
\begin{equation}
\label{eq:perm}
    P_{i} \, = \, S_i \cdot D_{i}.
\end{equation}

Permeability, a material property that is specific to a combination of membrane material and gaseous species, is measured in units of Barrer. The permeance $Q_{i}$ is used to evaluate a membrane rather than its material. $Q_i$ is the permeability $P_{i}$ per membrane thickness $\delta$, given in gas permeation units (GPU).
The selectivity $\alpha_{i,j}$ of a membrane, however, is given as the ratio of permeabilities or permeances, respectively. Hence, 
\begin{equation}
\label{eq:selectivity}
    \alpha_{ij} = \frac{P_{i}}{P_{j}} = \frac{Q_i}{Q_j}.
\end{equation}.

To fully evaluate its separation capacity, both the selectivity and the permeability of a membrane material need to be regarded. The combination of selectivity and permeability often shows a trade-off: polymeric membranes with good selectivity typically suffer from low permeability, while highly permeable membranes exhibit low selectivities. Lloyd Robeson \cite{robeson1991correlation} first visualized this trade-off in 1991 in his Robeson plots, introducing an upper bound of selectivity for certain gas pairs that was dependent on the permeation flux of the more permeable gas. He revisited this upper bound in 2008 and refined the mathematical relationship \cite{robeson2008upper}. In recent years, various FTMs have shown permeabilities and selectivities exceeding the Robeson upper limit for the \ce{CO2}/\ce{N2}, \ce{CO2}/\ce{CH4}, and \ce{CO2}/\ce{H2} gas pairs, respectivley. This makes FTM materials very interesting for, e.g., flue gas cleaning (\ce{CO2}/\ce{N2}), natural gas purification (\ce{CO2}/\ce{CH4}), or the water-gas shift reaction (\ce{CO2}/\ce{H2}), where \ce{CO2} has to be selectively removed from the system.

\subsection{Transport mechanisms of \ce{CO2} in FTMs}
\label{subsub:FTMechanism}
While all gaseous species are transported through the membrane according to the SDM, this conventional transport is superimposed by a carrier-mediated transport for \ce{CO2}. The carrier-mediated transport exploits chemical properties specific to \ce{CO2} and enhances its permeation through the membrane via reversible reaction mechanisms. The reversible reaction between a desired component A, in this case \ce{CO2}, and a carrier C is shown in the following Equation \ref{eq:reversible_carrier} \cite{rafiq2016role}:
\begin{equation}
\label{eq:reversible_carrier}
    \mathrm{Component}(A)+\mathrm{Carrier}(C) \ce{<=>}{AC}. 
\end{equation}.

\begin{figure}[ht!]
 \centering
 \includegraphics[width=0.9\linewidth]{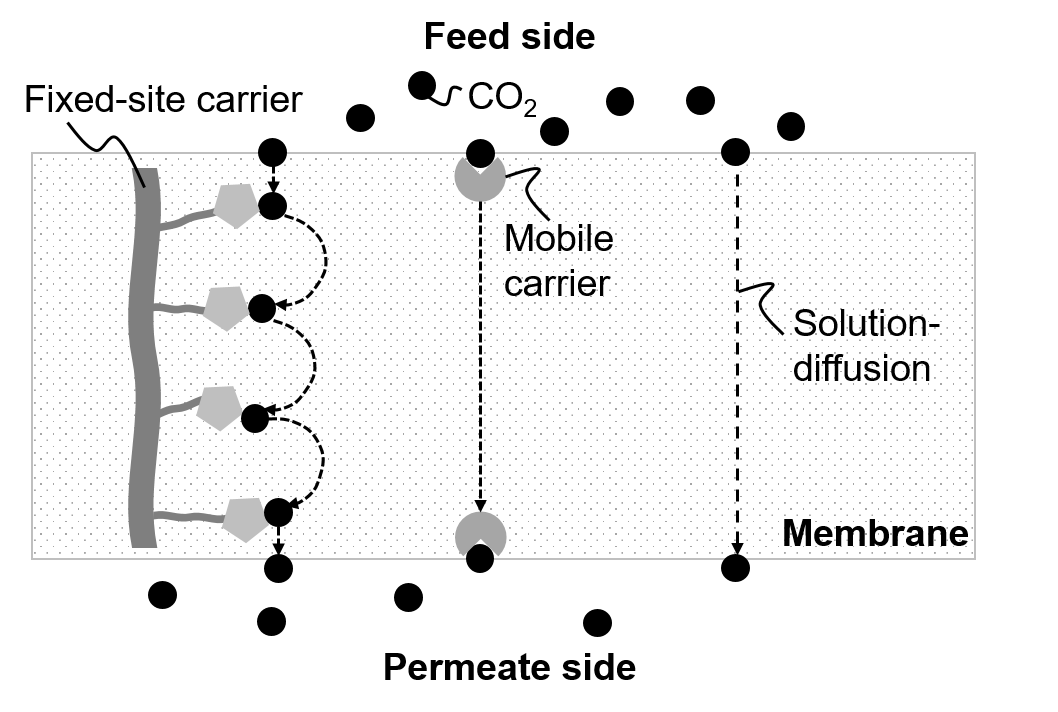}
 \caption{Schematics of transport mechanisms in facilitated transport membranes adapted from \cite{salim2018hydrogen}.}
 \label{fig:FTM}
\end{figure}

The overall flux of component $A$ ($J_{A}$)  is divided into the Fickian diffusion and the carrier-mediated diffusion, which therefore reads as 
\begin{equation}
\label{eq:flux_diffusion}
    J_A = \frac{D_A}{dx}(C_{A,f}-C_{A,p})+\frac{D_{AC}}{dx}(C_{AC,f}-C_{AC,p}), 
\end{equation}
where $D_A$ is the diffusion coefficient of the Fickian diffusion, and $D_{AC}$ is the diffusion coefficient of the carrier-mediated transport. Moreover, $C_A$ and $C_{AC}$ denote the concentration of component $A$ and the carrier solute $AC$ on the membrane interfaces, which are further denoted with $f$ for the feed side and $p$ for the permeate side, respectively \cite{rafiq2016role}. Figure \ref{fig:FTM} shows the carrier facilitated transport of \ce{CO2} through an FTM as it is commonly described in the literature \cite{salim2018hydrogen}.

Specific carrier materials interact with \ce{CO2} and reduce the transport resistance of the membrane. Two types of carriers prevail in current research:
\begin{itemize}
    \item{Fixed-site carriers: mostly polymers immobilized within the membrane}
    \item{Mobile carriers: smaller molecules dissolved within the membrane}
\end{itemize}
The reactive sites of the carriers in FTMs are usually amine groups. FTMs need humidity to function properly, because water and \ce{CO2} react within the membrane to form \ce{H2CO3}. The \ce{H2CO3} is subsequently deprotonated to form hydrogen carbonate (\ce{HCO3-}) and carbonate (\ce{CO3^2-}) in basic environments with pH level above 9, which are common when primary and secondary amines are present \cite{donaldson1980carbon, penny1983kinetic, versteeg1988kinetics}.

Within FTMs, two different transport models for the two carrier types, fixed-site and mobile, have been proposed. In membranes that host fixed-site carriers, the reaction product experiences a hopping mechanism along the charged amine groups \cite{cussler1989limits, zou2006co2}. It is similar to the Grotthuss mechanism introduced in 1805 \cite{von1805memoire}. The Grotthuss mechanism describes how protons are passed along water molecules, providing a theory for water conductivity long before the concept of ionized species was fully understood \cite{agmon1995grotthuss, cukierman2006tu, hassanali}. Recently, a comparable mechanism was introduced for ion transport through charged polymeric membranes, as ionized species hop along charged functional groups on the polymer backbone \cite{luduena11, luo2018selectivity}.

For mobile carriers, a coupled diffusion mechanism in which the carrier--\ce{CO2} adduct diffuses through the membrane along the concentration gradient of \ce{CO2} was proposed. After the transport through the membrane, the reaction between the carrier and \ce{CO2} is reversed, and \ce{CO2} exits the membrane in the permeate phase. \cite{zhao2013co2, tong2015water, maheswari2017separation}

\section{Materials and methods}
\label{ch:MnM}
\subsection{Materials}
Polyvinyl alcohol (PVA) \mbox{(MW=89-98~kDa)}, monoethanolamine (MEA) (purity $\geq$ 99\%), 4-amino-1-butanol (BA) (purity 98\%), and glutaraldehyde (GA) (25~wt.~\% in aq. solution, grade II) were purchased from Sigma-Aldrich (Merck). Microporous PTFE supports (BOLA, $\delta$~=~200~µm, $d_{p}$~=~0.2~µm) were purchased from Bohlender, whereas Polyethylene glycol 400 (PEG400) was ordered from Carl-Roth. Lupamin\textsuperscript{\textregistered} 9095, which was kindly donated by BASF, contains \mbox{12.5~wt.~\%} polyvinyl amine (PVAm) and about \mbox{9.5~wt.~\%} sodium formate, dissolved as \ce{Na+} and \ce{HCOO-} in water. Pure gases nitrogen (\ce{N2}) and carbon dioxide (\ce{CO2}) were ordered from Westfalen with a purity above \mbox{99.99~vol.~\%}. 

\subsection{Membrane preparation}
\label{ch:MembranePrep}
PVA was dissolved in deionized water at 80\,$^{\circ}\mathrm{C}$ and gently stirred for at least 4~h. Afterward, the temperature was reduced to 40\,$^{\circ}\mathrm{C}$, and GA was added as the chemical cross-linking agent at a mass ratio of 0.02~$\mathrm{m_{GA}}$/$\mathrm{m_{PVA}}$, which resulted in a cross-linking degree of 3.5\%. The solution was stirred overnight for at least 18~h. All carriers were added after the cross-linking of PVA to avoid reaction between carrier amine groups and the aldehyde groups of GA.

For the different membrane compositions, MEA and BA as mobile carriers and Lupamin\textsuperscript{\textregistered} containing PVAm as fixed-site carrier were added at 40\,$^{\circ}\mathrm{C}$. Table \ref{tb:Membranes} shows the different compositions, in mass fraction, that were tested within this work. The solutions were stirred for at least 1~h. The resulting membrane solutions were centrifuged at 8500~rpm for 10~min (Eppendorf 5804 centrifuge) to remove air bubbles, dust particles, and other impurities.

\begin{center}
		\begin{table}[htbp]
		\centering
			\begin{tabular} {l l l l l}
				\textbf{PVA} & \textbf{MEA} & \textbf{BA} & \textbf{PVAm} & \textbf{Sodium formate}\\
				\hline 
				1 & 0 & 0 & 0 & 0 \\
				0.9 & 0.1 & 0 & 0 & 0 \\
				0.858 & 0 & 0.142 & 0 & 0 \\
				0.765 & 0 & 0 & 0.135 & 0.1 \\
				0.725 & 0.1 & 0 & 0.1 & 0.075 \\
			\end{tabular}
		\caption{Overview of membrane synthesis solutions. Values represent mass fractions in dry membranes.}
		\label{tb:Membranes}
		\end{table}
\end{center}

After centrifuging, membranes were cast onto a glass plate with a casting knife film applicator (Elcometer~3580) with a gap setting of 800~µm. All membranes were cast from aqueous solutions with \mbox{10~wt.~\%} polymer content. The membranes were dried in a fume hood for at least 48~h at room temperature to evaporate any remaining solvent. Finally, the membranes were placed on microporous PTFE supports (BOLA) and were punched out with a diameter of 65~mm. All membranes were stored in enclosed containers at room temperature and a relative humidity of 50\%. The resulting dry membranes were 80~$\pm$~5~µm thick. 

\subsection{Sorption measurements}
\label{ch:sorptionmethod}
Ellipsometry is an optical measurement technique based on analyzing changes in the polarization of reflected light \cite{tompkins2005handbook, fujiwara2007spectroscopic}. Here, film thickness variations were measured with an ellipsometer (M-2000X, J.A. Woolam). The set-up used humidified gases \ce{CO2} and \ce{N2} to swell the samples and was operated at 25\,$^{\circ}\mathrm{C}$. For the eventual film thickness calculations, the Cauchy equation for transparent polymers was used \cite{ogieglo2012spectroscopic, ogieglo2015situ}. Samples were produced by spin coating (WS-650-23 NPP spin coater, Laurell) PVA on silica wafers with a thermally grown \ce{SiO2} layer. Dry PVA layers with a thickness of approximately 100~nm were synthesized from a \mbox{3~wt.~\%} aqueous PVA solution. The spin coating acceleration was set to 30~rpm/s with a constant maximum speed of 3000~rpm, which was held constant for 20~s. 

The weight gain in humid atmosphere was measured with an IGA moisture sorption analyzer (IGAsorp DVS analyzer, Hiden Isochema). The measurements were conducted with \ce{N2} at 25\,$^{\circ}\mathrm{C}$, as well as 50\,$^{\circ}\mathrm{C}$. 
Samples for the weight gain experiments were cast from an aqueous solution of \mbox{3~wt.~\%} PVA to a dry film thickness below 10~µm. 
Thickness variations and weight gain were measured with increasing humidity and consecutively with decreasing humidity to rule out hysteresis behavior. All samples were fully dried before measurements. Thicker PVA films for IGA measurements were dried at 100\,$^{\circ}\mathrm{C}$, while the thin ellipsometry samples were dried at ambient temperature in a dry \ce{N2} atmosphere.

\subsection{Permeation set-up}
\label{ch:ExpSetup}
\begin{figure}[ht!]
 \centering
 \includegraphics[width=0.9\linewidth]{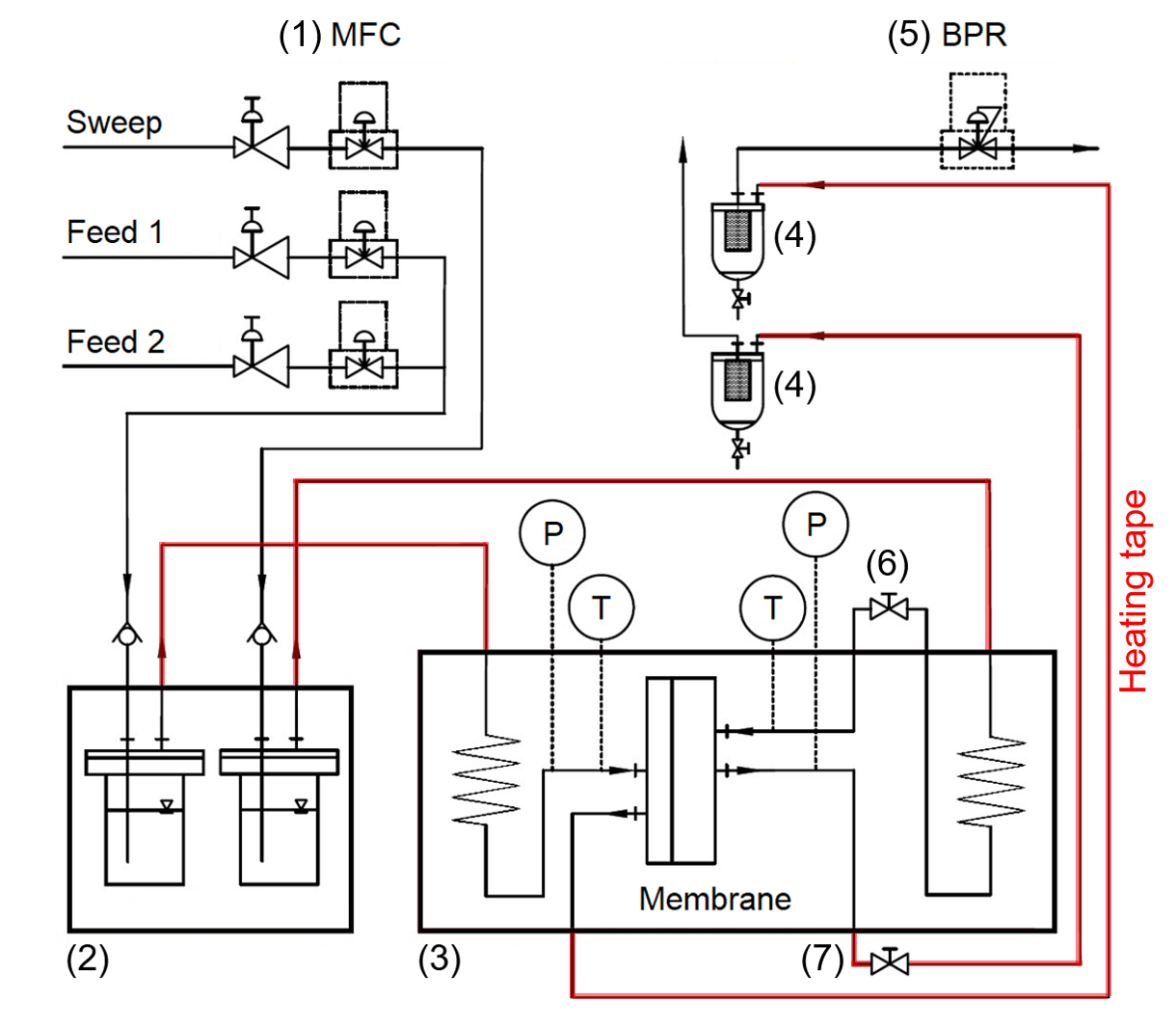}
 \caption{Simplified flow sheet of the testing apparatus: (1) mass flow controller (MFC), (2) heating bath for humidifiers, (3) heating bath for permeation cell, (4) water knockouts, (5) back pressure regulator (BPR), (6) sweep valve, (7) permeate valve. Red lines represent the installed heating tapes, which prevent condensation of \ce{H2O} in the pipes.}
 \label{fig:Flowsheet}
\end{figure}

Gas permeation measurements were conducted in a customary set-up, which is shown as a simplified flow sheet in Figure \ref{fig:Flowsheet}.
Inlets for gas streams on feed and sweep side were equipped with individual mass flow controllers (Analyt-MTC) (1). Gas streams were humidified by bubbling
through deionized water in stainless steel tanks before entering the permeation cell. The humidifiers were placed in a PEG400--water bath (2) and the temperature controlled by a thermostat (Lauda MT/2). Pipes connecting humidifiers, permeation cell, and water knockouts (4) were heated by a heating tape (SAF W\"armetechnik), which was temperature-controlled with an electrical heating system (Horst) to prevent condensation. The membrane cell was placed in a separate heating bath (3) filled with a PEG400--water mixture, also heated by a thermostat (Julabo FP 50). Separately controlled heating baths allowed for tuning of the relative humidity, since the saturation of the gaseous streams was decoupled from the membrane operation temperature. A back pressure regulator (BPR; Analyt-MTC) (5) was used to maintain a constant feed pressure throughout the experiments. After passing through the permeation cell, water knockouts forced the condensation of water. For pressure increase measurements, valves were placed on the sweep (6) and permeate side of the membrane (7), as described in Section \ref{ch:Perm1}

\subsection{Constant volume--variable pressure measurements}
\label{ch:Perm1}
Prior to measurements, the membranes were conditioned with humidified pure gas flowing on both the feed and sweep side at 50\,$^{\circ}\mathrm{C}$ and 60~mL/min, which swelled the membranes. After the conditioning period, the feed and sweep sides were purged of residual condensate or undesired species by a thorough flushing procedure. Immediately before the measurement, the volume flow on the sweep side was stopped and the valves for permeate and sweep connection were closed, with an absolute permeate pressure of 1~bar.

The pressure on the permeate side was monitored with a high-accuracy digital pressure indicator (Kobold) with a pressure range of 0--4~bar. The increased pressure over time was recorded manually, while the feed stream of 60~mL/min was kept constant at an absolute pressure of 2~bar and a relative humidity above 95\%. The temperature for the humidifiers and the permeation cell was constant at 50\,$^{\circ}\mathrm{C}$.

The resulting flux through the membrane was calculated according to Equation \ref{eq:Pinc}, introduced by Ballhorn \cite{ballhorn2000entwicklung}, which is based on the compression behavior of ideal gases. The average pressure difference during permeation was calculated as the logarithmic mean value over time.

\begin{equation}
\label{eq:Pinc}
    J = \frac{ V_{p} \cdot V_{mol} } {A \cdot R \cdot T \cdot t_{1} } \cdot ln \left( \frac{ p_{F} - p_{P} ( t_{0} )}{ p_{F} - p_{P} ( t_{1} )}\right)
\end{equation}.

\begin{equation}
\label{eq:Vmolar}
    V_{mol} = V_{mol,STP} \cdot \frac{T}{273.15\,K}
\end{equation}.

In Equation \ref{eq:Pinc}, $J$ represents the flux through the membrane, $V_{p}$ is the enclosed volume on the permeate side, $A$ is the membrane area, and $R$ is the universal gas constant. $p_{F}$ is the feed pressure, which was held constant over permeation time. $p_{P}(t_0)$ is the permeate pressure at the beginning of the measurement, while $p_P(t_1)$ is the permeate pressure at the end of the measurement. $V_{mol}$ is the molar volume of a gaseous species at temperature T, calculated as depicted in Equation \ref{eq:Vmolar} according to the law for ideal gases, with \mbox{$V_{mol,STP}$~=~24.4~mol/L}.

\ce{CO2} pure gas measurements were repeated at least three times. Between every cycle, feed and sweep side were flushed for 10~min to ensure constant humidification of the membrane. All measurements were conducted with three different membranes, cast from the same solution. Furthermore, \ce{N2} pure gas measurements were conducted with each membrane, to check for defects.

\subsection{Computational methods}
\label{sec:compdet}

\begin{figure}[ht!]
    \centering
    \includegraphics[width=0.2\textwidth]{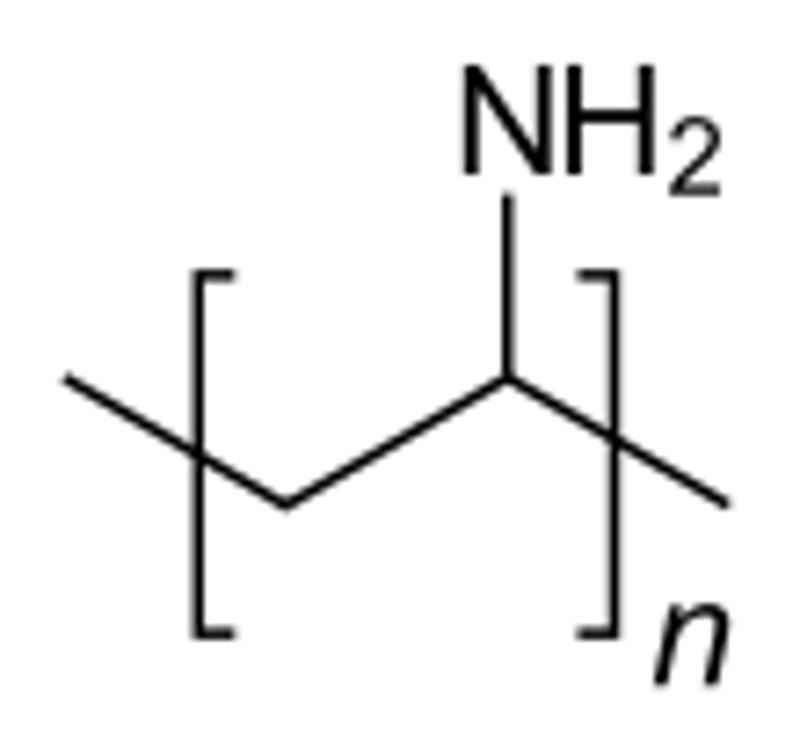}
    \caption{Structural formula of polyvinyl amine (PVAm).}
    \label{fig:PVAm_struct}
\end{figure}

\begin{figure}[ht!]
    \centering
    \includegraphics[width=0.49\textwidth]{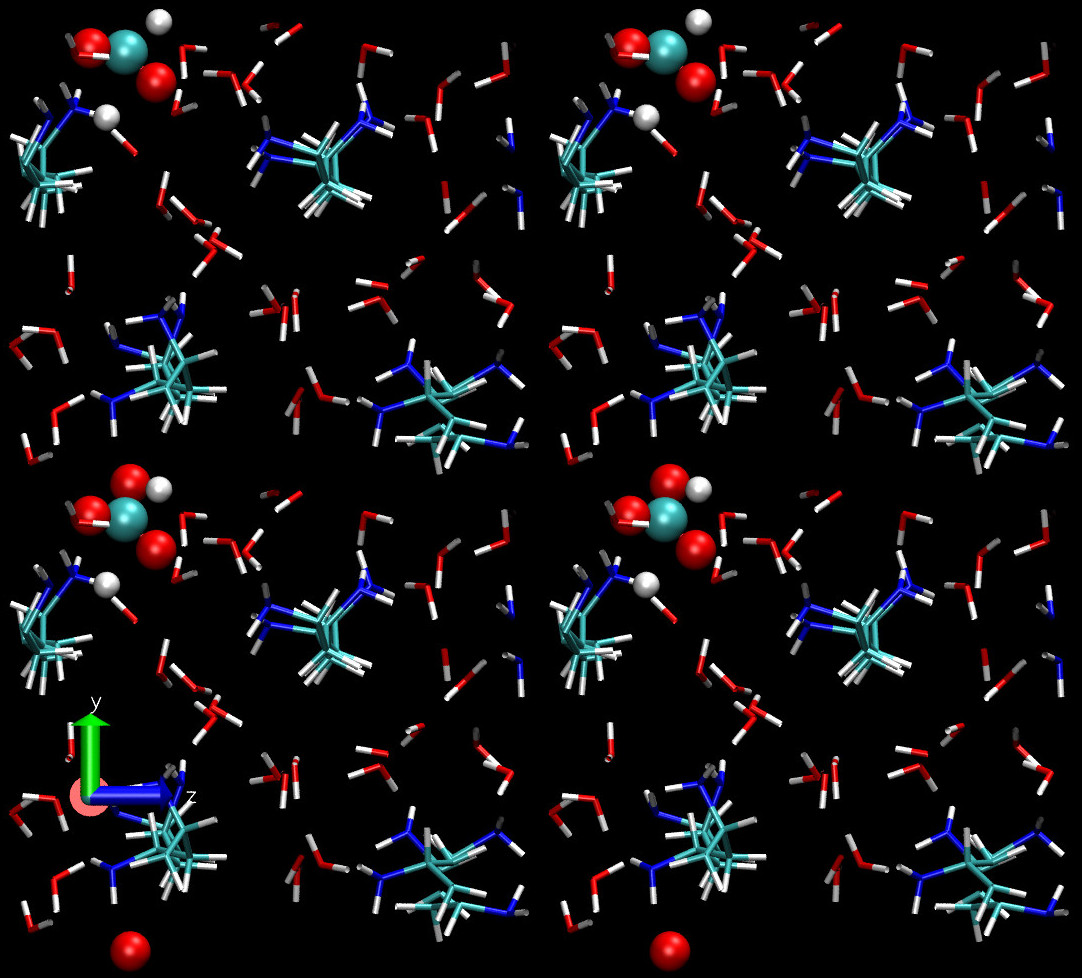}
    \includegraphics[width=0.49\textwidth]{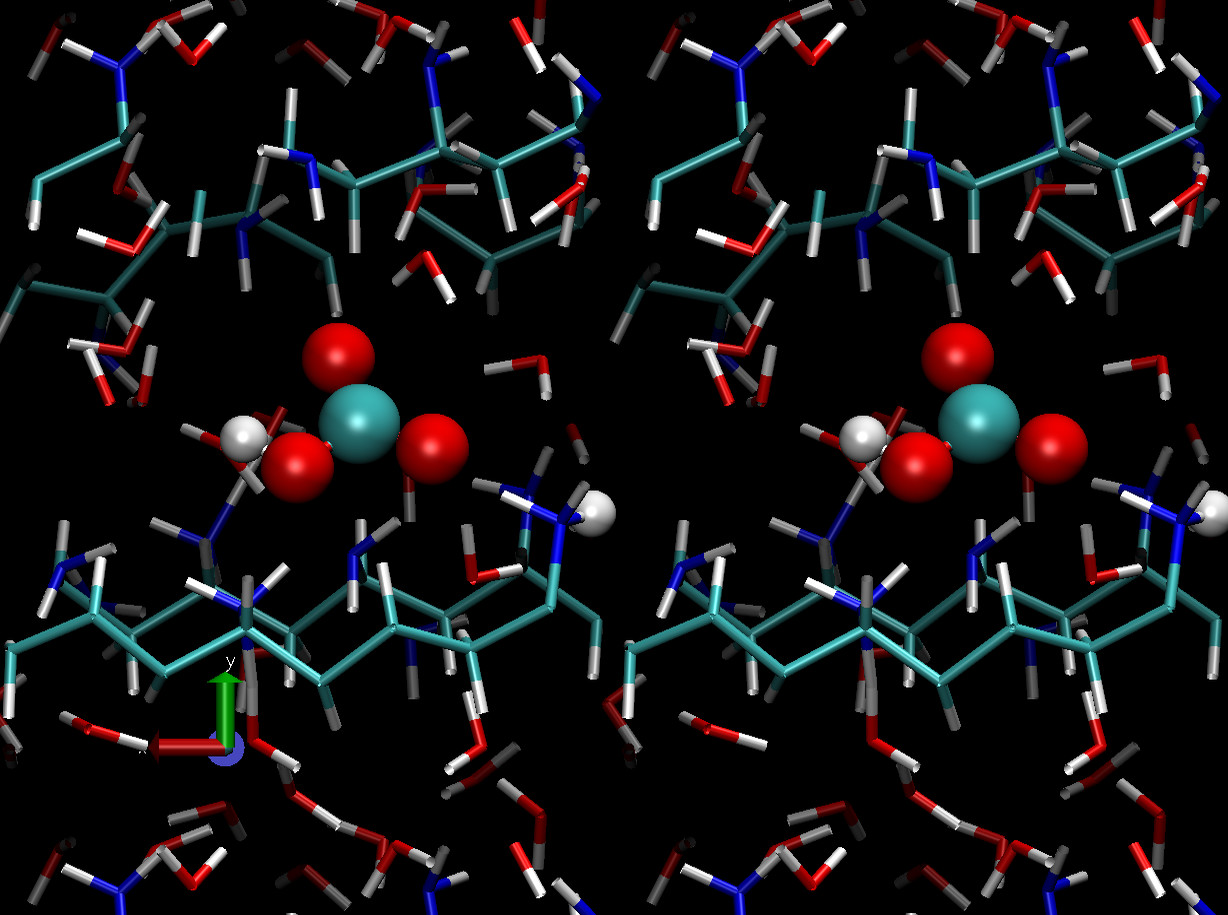}
    \caption{Snapshots of a simulated system along the z-axis (left panel) and x-axis (right panel), respectively. The bonds of the polymer chains and water molecules are depicted by lines, whereas the atoms of \ce{H2CO3} are represented with spheres. Carbon atoms are shown in cyan, nitrogen atoms in blue, oxygen atoms in red, and hydrogen atoms in white.}
    \label{fig:sideview}
\end{figure}

Using the model based on our previous work \cite{luduena11}, we simulated systems with 4 independent linear syndiotactic polyvinyl amine (PVAm, structural formula shown in Figure \ref{fig:PVAm_struct}) chains in a hexagonal arrangement with distances of 8~\AA\ between the chains and different concentrations of \ce{H2CO3} in water solution. Representative snapshots of one of the simulated systems can be seen in Figure \ref{fig:sideview}. The chains are oriented along the x-axis of the periodic simulation cell and therefore correspond to an infinitely extended polymer. Each chain consists of 8 carbon atoms and 4 amino groups, which results in 16 amino groups per unit cell, whose dimensions are 10.25~\AA, 13.86~\AA, and 16~\AA\ along the x, y, and z directions, respectively. 

Specifically, we studied three different systems: a dry polymer without water, as well as a "semi-dry" with 18 water molecules (1.125 \ce{H2O} per amino group) and a "wet" polymer with 36 water molecules (2.25 \ce{H2O} per amino group) in the unit cell. The initial structure and locations of the polymer chains were prepared using the software Avogadro \cite{avogadro,avogadropaper}, whereas the water molecules were placed at random between the chains using the software Packmol \cite{packmol}.

Our density functional theory-based AIMD simulations were conducted using the Gaussian and plane waves-based second-generation Car-Parrinello method of Kühne and coworkers \cite{kuhne2007efficient, CP2G, Prodan, HUTTER2023}, as implemented in the CP2K program package \cite{cp2k_jcp}. Therein, the Kohn-Sham orbitals were expanded in contracted Gaussians, whereas the electronic charge density was represented using plane waves. For the former, we employed a molecularly optimized triple-$\zeta$ basis set with one additional set of polarization functions (TZVP) \cite{molopt}, whereas a density cutoff of 280~Ry was used for the latter. The core electrons were represented by norm-conserving Goedecker-Teter-Hutter (GTH) pseudopotentials \cite{GTH1,GTH2,GTH3}, and the unknown exchange-correlation potential was substituted by the B3LYP hybrid exchange-correlation functional \cite{BLYP1,BLYP2}.

For each system, first a 1~ps long unbiased NVT equilibration run at 400~K was performed, with a discretized time step of 0.25~fs. Subsequently, now using a time step of 0.5~fs, 50~ps long NVT production runs were conducted, where a force bias along the x-axis was applied to the atoms of \ce{HCO3-}. In all of our simulations, the temperature was controlled by means of the CSVR thermostat at a time constant of 50~fs \cite{CSVR}. 

In our simulations, we find that \ce{H2CO3} rapidly deprotonates and forms a \ce{R-NH3+HCO3-} ion pair with the nearest amino (\ce{NH2}) group, which is in line with the results of a previous study by Daschakraborty et al. \cite{daschakraborty16}. 
An external force bias along the polymer chains was applied to the atoms of the resulting hydrogen carbonate ion to accelerate the transport along the polymer chains. The magnitude of the force bias of 90~µN was chosen such that there was separation of the \ce{R-NH3+HCO3-} ion pair. The distance traveled by the \ce{HCO3-} anion relative to the \ce{R-NH3+} ion, along the direction of the bias force, was computed to monitor the transport.

\section{Results and discussion}
\label{ch:RnD}

\subsection{Sorption measurements}
\label{ch:sorption}
\begin{figure}[ht!]
 \centering
 \includegraphics[width=0.7\linewidth]{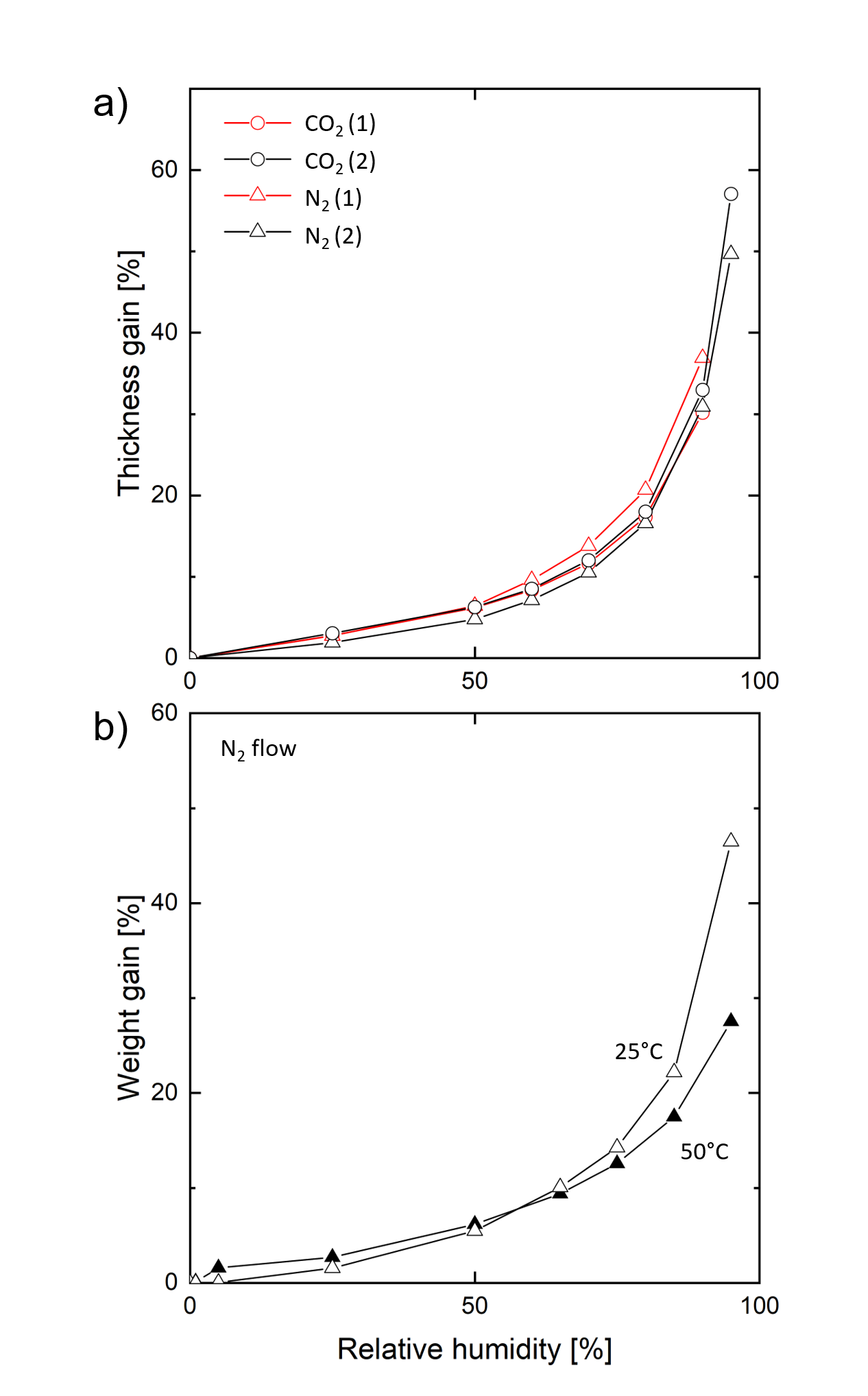}
 \caption{Sorption measurements were conducted in PVA films. a) Relative thickness gain measured by ellipsometry in \ce{N2} and \ce{CO2} atmospheres at 25\,${\circ}\mathrm{C}$ with PVA films of 100~nm thickness. b) Relative weight gain assessed with IGA sorption measurements, which were conducted in \ce{N2} atmosphere at 25\,${\circ}\mathrm{C}$ and 50\,${\circ}\mathrm{C}$ with PVA films $\leq$10~µm thick.}
 \label{fig:sorption}
\end{figure}

Sorption measurements were conducted as described in Section \ref{ch:sorptionmethod}. The ellipsometry measurements assessed swelling by measuring changes in the relative film thickness, as shown in Figure \ref{fig:sorption}a. IGA sorption measurements assessed sorption through a relative weight gain, as presented in Figure \ref{fig:sorption}b. Both measurements show an exponential increase of swelling with increased humidity. The increase implies a transition from the glassy to the rubbery polymer state for PVA at higher humidity levels, with water as the plasticizer. This transition is in accordance with the literature, where penetrant-induced plasticization has been well studied \cite{van1999sorption, ogieglo2014polymer, ogieglo2014effective}. The desorption of water from PVA showed no measurable hysteresis effect.

\ce{N2} acts as an inert gas with low absorption in water, while \ce{CO2} dissolves in water as \ce{H2CO3}. Therefore, a decreased pH within the swollen membrane under a \ce{CO2} atmosphere was expected. Research conducted on the sorption properties of PVA films cast from solutions of different pH showed decreased swelling at lower pH values \cite{mansur2008ftir}. However, no apparent impact on thickness in either \ce{CO2} or \ce{N2} atmosphere was detected, as Figure \ref{fig:sorption}a shows. The weak acidity of \ce{H2CO3}, with an apparent acid dissociation constant (pK(app)) of 6.35, seems to be insufficient to measurably affect the swelling behavior of PVA.

The results of the IGA sorption measurements from Figure \ref{fig:sorption}b show a maximum weight gain of 47\% at 25\,$^{\circ}\mathrm{C}$ and 27\% at 50\,$^{\circ}\mathrm{C}$ for 95\% relative humidity. Comparing the degrees of swelling at both temperatures, a decreased water uptake at higher temperatures was observed. The relative weight gains are in agreement with literature values \cite{figueiredo2009poly}. The permeation measurements at elevated temperatures highlight the need to work at close to 100\% relative humidity in order to supply a sufficient amount of water for the FTM to work.

\subsection{Permeation measurements}
\label{sec:permeation_measure}
Permeation measurements were conducted according to Section \ref{ch:Perm1}. The permeation measurements were assessed by permeance rather than permeability because of the membrane swelling behavior in humid atmosphere and the unknown membrane thickness after swelling. Repeat measurements of \ce{CO2} permeance for each membrane showed good reproducibility with deviations below 10\%. \ce{N2} permeation experiments were performed overnight, to check for any defects in the membrane. Dry PVA flat sheet membranes did not show any measurable permeance for \ce{N2} or \ce{CO2}. These results were in accordance with the literature, as dry PVA acts as a gas barrier \cite{pye1976measurement, salame1977barrier}. However, Vega et al. \cite{aguilar1993gas} observed that PVA swells when exposed to liquid water or humid atmospheres, which increases its permeability by up to three orders of magnitude. In addition, humidified PVA has a \ce{CO2}/\ce{N2} selectivity, which occurs because of the relatively high \ce{CO2} solubility in water compared to the solubility of \ce{N2} \cite{linstrom2005nist}. The results from Section \ref{ch:sorption} confirm the swelling of PVA in high relative humidities. Therefore, all ensuing measurements were conducted with relative humidities above 95\% to ensure sufficient membrane swelling.

\subsubsection{Mobile carrier-induced permeation through FTMs}
\begin{figure}[ht!]
 \centering
 \includegraphics[width=0.9\linewidth]{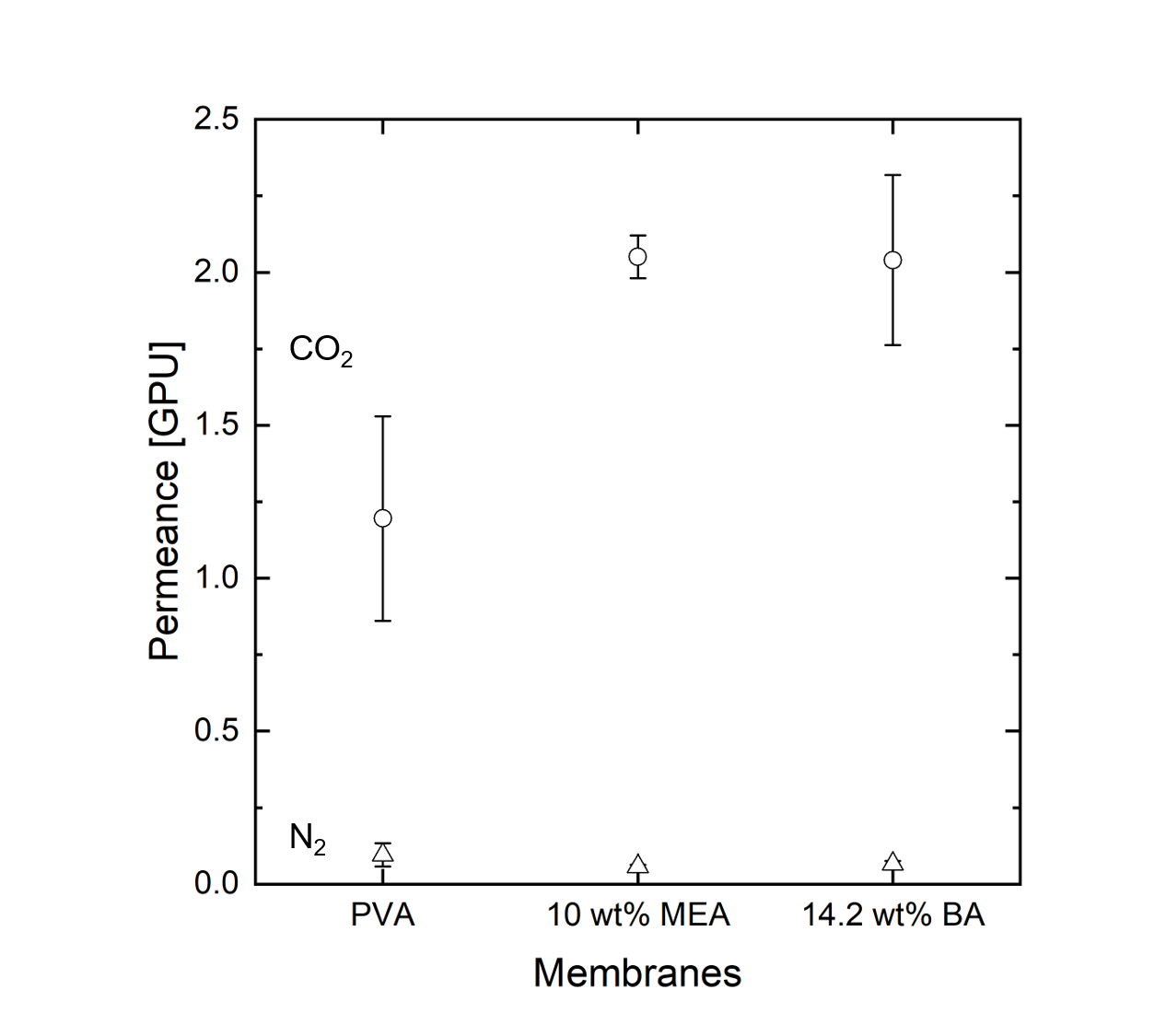}
 \caption{Permeation measurements of \ce{CO2} through PVA, PVA + \mbox{10~wt.~\%} MEA, and PVA + \mbox{14.2~wt.~\%} BA membranes. All measurements were conducted with 1~bar TMP at 50\,${\circ}\mathrm{C}$ and a relative humidity above 95\%.}
 \label{fig:PVA_mobiles}
\end{figure}

MEA and BA were used as mobile carriers to assess the permeation characteristics of carrier molecules of different sizes. MEA and BA have a similar molecular structure, with one primary amine and one hydroxyl group per molecule. The difference is the molecular mass of BA (89.14 g/mol) compared to MEA (61.08~g/mol), with a ratio of ${M_{BA}}/{M_{MEA}}$~=~1.46. The membranes were prepared with the same molar concentration of either amine compound to allow a comparison between the two mobile carriers. Therefore, membranes containing 10~wt.~\% MEA and 14.2~wt.~\% BA were compared to pure PVA membranes, and the results are displayed in Figure \ref{fig:PVA_mobiles}.

Pure PVA showed a permeance of 1.2~GPU for \ce{CO2}. The addition of either mobile carrier, MEA or BA, nearly doubled the permeance, while the \ce{N2} permeance remained constant and very low in both cases. With a higher molecular weight, BA has a lower diffusion coefficient than MEA \cite{cohen1959molecular}. If carrier-facilitated transport were solely dependent on diffusive transport of the carrier--\ce{CO2} adduct, a lower permeance for the membrane featuring BA as a carrier would always be expected. However, as shown in Figure \ref{fig:PVA_mobiles}, both membranes show a similar permeance for \ce{CO2}. These results indicate that the diffusion of adducts is not the only transport mechanism for FTMs hosting mobile carriers. In addition to the diffusion, a hopping mechanism for \ce{CO2}, dissolved as a hydrogen carbonate (\ce{HCO3-}) or carbonate (\ce{CO3^2-}), between charged molecules within the membrane seems plausible.

\subsubsection{Fixed-site and mixed carrier-induced permeation through FTMs}
\begin{figure}[ht!]
 \centering
 \includegraphics[width=0.9\linewidth]{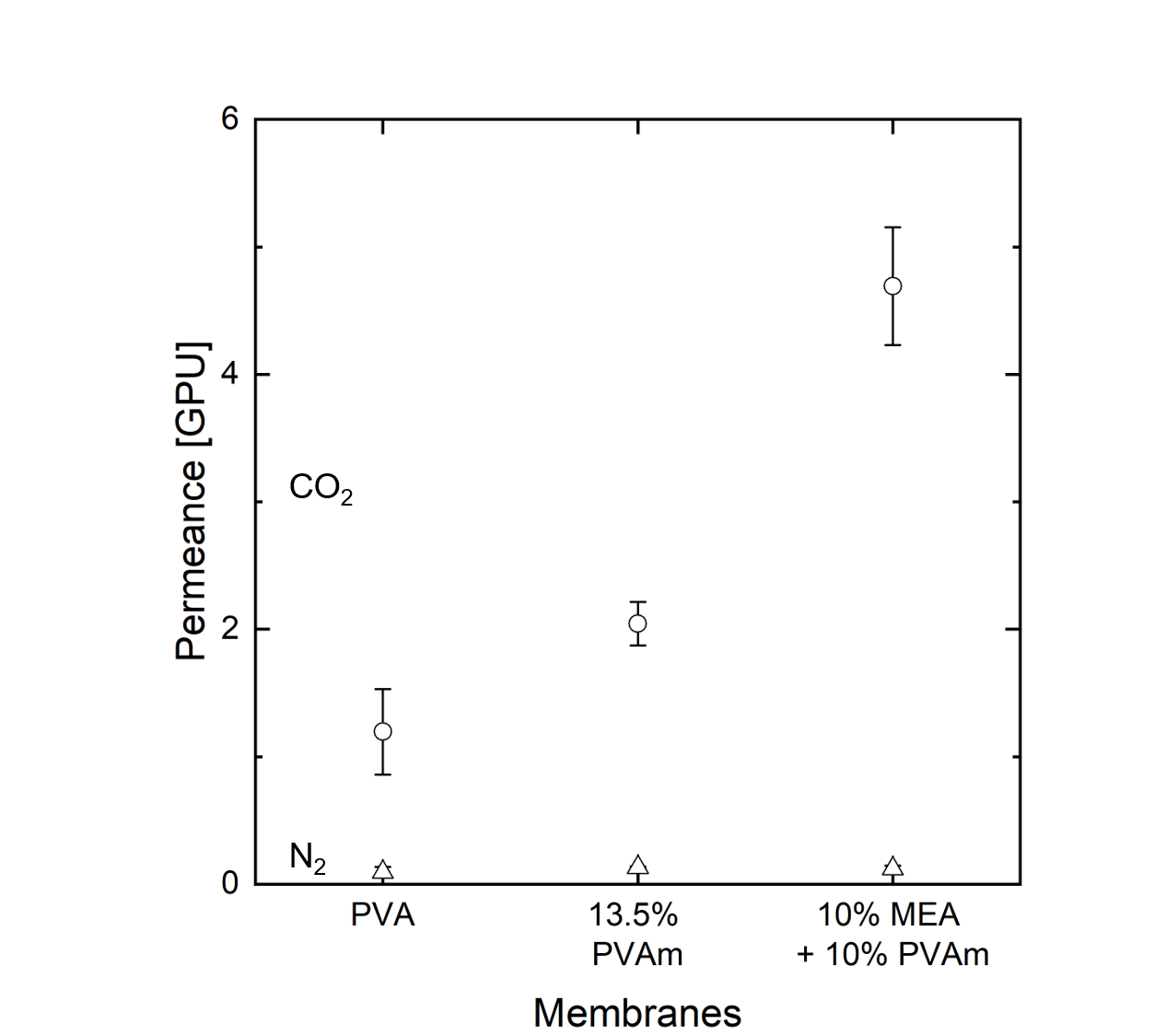}
 \caption{Permeation measurements of \ce{CO2} through PVA, PVA + \mbox{13.5~wt.~\%} PVAm, and PVA + \mbox{10~wt.~\%} PVAm + \mbox{10~wt.~\%} MEA membranes. All measurements were conducted with 1~bar TMP at 50\,${\circ}\mathrm{C}$ and a relative humidity above 95\%.}
 \label{fig:PVA_mixed}
\end{figure}

The results from permeation measurements with membranes featuring different carriers are displayed in Figure \ref{fig:PVA_mixed}. The membranes were synthesized with Lupamin\textsuperscript{\textregistered} PVAm from BASF.

The increased \ce{CO2} permeance through membranes with PVAm as fixed-site carrier compared with pure PVA indicate a carrier-facilitated transport. The addition of \mbox{13.5~wt.~\% PVAm}, with one primary amine group per repeating unit of 43.07~g/mol molecular mass, added more than twice the amount of functional amine groups to the membrane than \mbox{10~wt.~\%} MEA or \mbox{14.2~wt.~\%} BA. Surprisingly, the permeances of the fixed-site carrier membranes remained on the same level as the mobile carrier permeances. This effect could be explained by the movement of mobile carriers within the membrane structure according to the chemical potential. MEA and BA can freely change position within the polymer for optimal amine distribution, whereas PVAm is bound to its polymeric chain. Therefore, not all amine groups of PVAm are equally contributing to the \ce{CO2} transport, in contrast to the mobile carriers. 

Interestingly, for mixed-carrier membranes with fixed-site carrier PVAm and mobile carrier MEA, a drastic increase in permeance for \ce{CO2} to 4.7~GPU was measured. Compared with the measurements with solely mobile or fixed-site carriers, the permeance more than doubled. At the same time, the \ce{N2} permeance remained low and no defects were measurable in the membranes. The increase in permeance for the presented mixed-carrier membrane was larger than the individual increases attributed to either carrier. These findings are in agreement with those in the literature, where high permeances were observed when combining fixed-site and mobile carriers \cite{zou2006co2, zhao2013co2, qiao2015preparation}. The results imply a synergistic effect of different-type carriers within the FTM.

These synergies could be due to differences in localization and mobility of mobile and fixed-site carriers within the membrane matrix. While PVAm is immobilized by its polymeric chain length, MEA molecules remain mobile, especially when the PVA membrane matrix is swollen with water. Without fixed-site carriers, a uniform concentration of MEA in the membrane is assumed. However, the high density of generally positively charged amine groups in PVAm most likely interact with the charged amine groups of MEA. Thus, MEA occupies spaces with lower PVAm concentration, enabling facilitated transport through areas with lower local PVAm concentration.

Uncertainty remained regarding the role of water molecules in the transport through the membrane. Transport through FTMs without water is slow and unselective. Therefore, AIMD simulations, described in Section \ref{sec:compdet}, were used to understand the interaction of \ce{CO2} with amine groups and \ce{H2O} molecules, and the results are presented in Section \ref{sec:simres}. 

\subsection{Discussion of the \textit{ab initio} simulations}
\label{sec:simres}

Within the first several hundred femtoseconds of the unbiased equilibration run, \ce{H2CO3} was immediately deprotonated, forming \ce{HCO3-}. The nearby amino group captured the proton from the carbonic acid, thus forming a protonated amino group \ce{R-NH3+}. This observation, which is consistent with calculations by Daschakraborty et al. \cite{daschakraborty16}, was expected, since \ce{H2CO3} is a very unstable species that readily reacts with any Brønsted-Lowry base. 

During our production runs, \ce{HCO3-} was pulled along the x-axis with a force bias of 90~µN. This specific value proved to be strong enough to separate the \emph{initial} \ce{R-NH3+ HCO3-} ion pair yet weak enough to produce a slow steady movement of \ce{HCO3-} of approximately 3~\AA/ps relative to the polymer backbone. In the right panel of Figure \ref{fig:sideview}, it can be seen that the protonated amino group and \ce{HCO3-} anion are quite far apart.

The proton captured by the amino group remained attached for the whole duration of the production run, which means that the initial proton hopping in the equilibration run was the only proton-hopping event observed. In an additional simulation run, we applied the force bias also to the exchanged proton, but the N--H bond did not break, even up to a magnitude of 300~µN.

\begin{figure}[ht!]
    \centering
    \includegraphics[width=0.8\textwidth,clip,trim={0 0 0 0cm}]{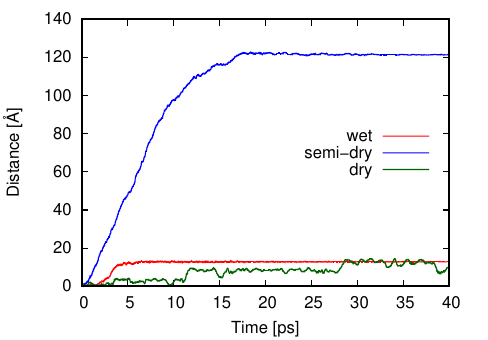}
    \caption{Distance traveled along the x-axis by \ce{HCO3-} relative to the protonated amino group as a function of time.}
    \label{fig:distvst}
\end{figure}

The distance traveled by \ce{HCO3-} relative to the protonated amino group as a function of time is plotted in Figure \ref{fig:distvst}. In the wet and dry cases, the distance is approximately 10~\AA, which roughly corresponds to the length of the simulation box in the x direction (10.25~\AA). In the semi-dry case, however, the \ce{HCO3-} anion traverses a much longer distance than in the other two cases, of roughly 120~\AA. That is to say, \ce{HCO3-} traveled through the box 12 times before stopping. Also, the slope of the semi-dry system is steeper than in the case of the wet system, which means that the velocity of \ce{HCO3-} in the former system is much higher and thus the transport faster. 

Since both, the distance covered by \ce{HCO3-} and the velocity of the ion relative to the polymer backbone is higher in the case of the semi-dry polymer, it can be concluded that the transport across such a polymer is faster and less hindered than in the other two cases. This immediately suggests an underlying water-mediated Grotthus-mechanism, as first suggested by Ludueña et al.~\cite{luduena11}. After some time, the curves of Figure \ref{fig:distvst} level off, which indicates that from a certain point on, the distance traveled by \ce{HCO3-} relative to the protonated amino group stops increasing. In other words, since the box is periodic, \ce{HCO3-} will eventually reach the protonated amino group again and the \ce{R-NH3+HCO3-} ion pair can re-form.

\subsection{Combining the experimental and the AIMD simulations findings}
\begin{figure}[ht!]
 \centering
 \includegraphics[width=0.9\linewidth]{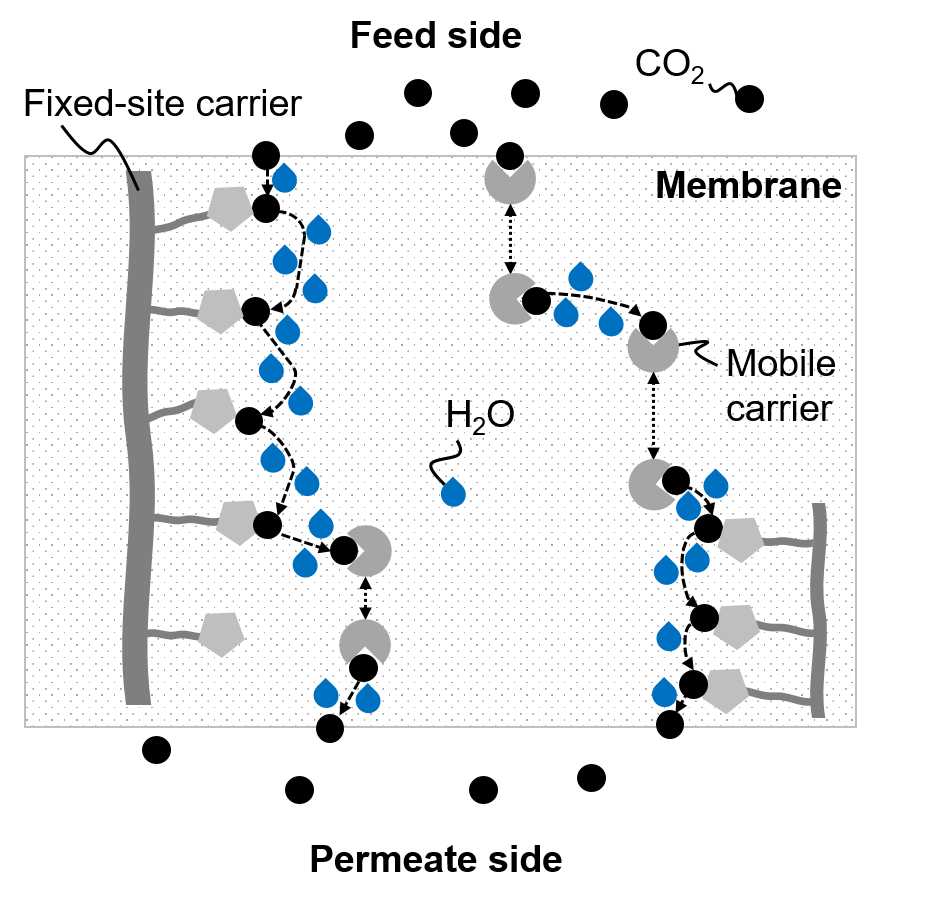}
 \caption{Schematics of a proposed combined hopping mechanism, with water as an intermediate charge carrier. The black circle denotes \ce{CO2} and its different forms during the transport, e.g., \ce{H2CO3}, \ce{HCO3-}, and \ce{CO3^2-}.}
 \label{fig:combined_hopping}
\end{figure}

Based on the findings from Section \ref{sec:permeation_measure} and Section \ref{sec:simres}, a combined hopping mechanism of \ce{CO2}, between charged amino groups of mobile and fixed-site carriers, is assumed in the presence of water molecules, as shown in Figure \ref{fig:combined_hopping}. This Grotthus-like hopping mechanism needs water to form \ce{H2CO3}, and its dissociation generates \ce{HCO3-} and, in a sufficiently alkaline environment \ce{CO3^2-}, which are then transported through the membrane as charged ions. Water is assumed to hydrate \ce{HCO3-} once it gets in close vicinity, which lowers the attraction between \ce{R-NH3+} and \ce{HCO3-}. This can cause hopping from one amine carrier to the next. In addition, the experiments showed that water swells the membrane matrix PVA and enhances the solubility of \ce{CO2} in the membrane, which is necessary for fast transport through the membrane. 
 
\section{Conclusion}
In this study, the transport mechanisms of \ce{CO2} through FTMs were investigated based on experiments and AIMD simulations. During the experiments, PVA was used as membrane matrix material, with PVAm as a fixed-site carrier and/or either MEA or BA as a mobile carrier. \ce{CO2} transport through pure PVA membranes in a dry state showed no measurable permeate gas flux. Once humidified, the PVA material swelled and the \ce{CO2} flux through the membrane enhanced, even though PVA does not contain any amine groups. Both results fit well with the literature, where dry PVA is presented as a gas barrier whereas humidified PVA is known to swell and show enhanced gas transport.

For a comparison of mobile carriers, measurements with MEA and BA were conducted. Based on literature, it was assumed that the \ce{CO2} transport with the same mole ratio of amine would be faster in the MEA/PVA membrane compared to the BA/PVA because of the smaller molecular size of MEA, which would benefit the diffusion of the MEA/\ce{CO2} complex. However, even though the permeate flux was enhanced for both mobile carriers compared to pure PVA, it remained at comparable levels between the two cases.

In a subsequent experiment, PVAm was integrated into PVA as a fixed-site carrier. For the fixed-site \ce{CO2} transport, a hopping mechanism between the amine groups of PVAm and \ce{CO2} in the form of \ce{H2CO3} or a dissociated form, e.g., \ce{HCO3-}, was expected. The experimental results agreed with expectation, as the \ce{CO2} permeate flux increased to similar values as in the PVA/MEA membrane.

In the last experiment, fixed-site and mobile carriers were combined to form a PVA/PVAm/MEA membrane. Interestingly, the permeate flux exceeded the sum of the respective individual permeances. This apparent superposition of permeances led to the idea of a combined \ce{CO2} transport mechanism, mixing carrier diffusion of the mobile carriers with the hopping mechanism of \ce{CO2} and fixed-site carriers.

For the purpose to elucidate the influence of water molecules on the transport rate, AIMD simulations were conducted to investigate the fixed-site carrier mechanism based on the transport of \ce{HCO3-} along PVAm polymer chains with different water contents. These simulations of polymers with different water content suggested a water-mediated Grotthus-mechanism, wherein water is vitally important for a hopping mechanism to take place in the case of fixed-site polyvinyl amine carriers.
Furthermore, solvation of \ce{HCO3-} reduces the ion-ion interaction, thus allowing the \ce{R-NH3+HCO3-} ion pair to separate and allowing the \ce{HCO3-} anion to hop to another protonated amino group. It also turned out that after a certain fraction of water in a polymer is reached, the transport becomes slower because of higher hydration of \ce{HCO3-}, which in turn increases its hydrodynamic radius and reduces its mobility. 

We conclude by noting that a future theoretical investigation of the interaction between a chosen mobile carrier and the \ce{R-NH3+HCO3-} ion pair, or perhaps even the \ce{R-NH3+CO3^2-} ion pair, would allow to further elucidate the transport mechanism in the presence of both fixed-site and mobile carriers. Using a similar simulation protocol, it could be determined whether the mobile carrier can separate the \ce{R-NH3+HCO3-} ion pair and thus accelerate the transport via the proposed hopping mechanism. If this turned out to be the case, it would account for the synergistic effect observed in the system with fixed-site carriers and mobile carriers present at the same time.

\section*{Acknowledgement} 
 The authors gratefully acknowledge financial support from the European Commission within the Horizon2020-SPIRE project ROMEO (grant agreement number 680395). Part of the project has received funding from the European Research Council (ERC) under the European Union’s Horizon 2020 research and innovation program (grant agreement no. 716142, as well as the Collaborative Research Centre SFB-TR87 of the German Science Foundation (DFG). 
 The generous allocation of computing time on the FPGA-based supercomputer ``Noctua'' by the Paderborn Center for Parallel Computing (PC$^2$) is kindly acknowledged.

\bibliographystyle{elsarticle-num}
\bibliography{FTM_MD}{}

\end{document}